# Functioning of the dimeric GABA$_B$ receptor extracellular domain revealed by glycan wedge scanning

Philippe Rondard[1,§], Siluo Huang[2,§], Carine Monnier[1], Haijun Tu[2], Bertrand Blanchard[1], Nadia Oueslati[1], Fanny Malhaire[1], Ying Li[2], Eric Trinquet[3], Gilles Labesse[4,5], Jean-Philippe Pin[1] & Jianfeng Liu[2]

[1] CNRS, UMR 5203, Institut de Génomique Fonctionnelle, Montpellier, France and INSERM, U661, Montpellier, France and Université Montpellier 1,2, Montpellier F-34000, France.

[2] Sino-France Laboratory for Drug Screening, Key Laboratory of Molecular Biophysics of Ministry of Education, Huazhong University of Science and Technology, Wuhan, Hubei, China

[3] CisBio International, Parc technologique Marcel Boiteux, Bagnols/Cèze cedex F-30204, France

[4] Centre de Biochimie Structurale, CNRS UMR5048, Université Montpellier 1, Montpellier F-34060, France

[5] INSERM U414, Montpellier, F-34094 France

[§] P.R. and S.H. contributed equally to this work

Correspondence should be addressed to J.P.P (jppin@igf.cnrs.fr) and J.L. (jfliu@mail.hust.edu.cn).

**Running title :** Functional analysis of GABA$_B$ receptor VFTs using N-glycans

Total character count (including spaces) : 55,058

**Abstract**

The G-protein coupled receptor activated by the neurotransmitter GABA is made up of two subunits, $GABA_{B1}$ and $GABA_{B2}$. While $GABA_{B1}$ binds agonists, $GABA_{B2}$ is required for trafficking $GABA_{B1}$ to the cell surface, increasing agonist affinity to $GABA_{B1}$, and activating associated G-proteins. These subunits each comprise two domains, a Venus flytrap (VFT) domain and a heptahelical (7TM) domain. How agonist binding to the $GABA_{B1}$ VFT leads to $GABA_{B2}$ 7TM activation remains unknown. Here, we used a glycan wedge scanning approach to investigate how the $GABA_B$ VFT dimer controls receptor activity. We first identified the dimerization interface using a bioinformatics approach, and then showed that introducing an N-glycan at this interface prevents the association of the two subunits and abolishes all activities of $GABA_{B2}$, including agonist activation of the G-protein. We also identified a second region in the VFT into which insertion of an N-glycan does not prevent dimerization, but does block agonist activation of the receptor. These data provide new insight into the function of this prototypical GPCR, and demonstrate that a change in the dimerization interface is required for receptor activation.

## Introduction

GABA is the main inhibitory neurotransmitter in the central nervous system, regulating many physiological and psychological processes. It mediates fast synaptic inhibition through ionotropic $GABA_A$ receptors, as well as slow and prolonged synaptic inhibition through both pre- and post-synaptic metabotropic $GABA_B$ receptors (Bettler and Tiao, 2006; Couve et al., 2004). $GABA_B$ receptors represent promising drug targets for the treatment of epilepsy, pain, drug addiction, anxiety, and depression (Bowery, 2006; Cryan and Kaupmann, 2005). $GABA_B$ agonists have demonstrated beneficial effects in humans, as illustrated by the anti-spastic activity of baclofen (Lioresal®). Recently, $GABA_B$-positive allosteric modulators (PAM) have been identified as potentially better alternatives to agonists, as they can limit the development of tolerance and avoid the adverse effects observed with agonists (Pin and Prézeau, 2007).

At the structural level, the $GABA_B$ receptor is composed of two homologous subunits, $GABA_{B1}$ and $GABA_{B2}$ (Jones et al., 1998; Kaupmann et al., 1998; Kuner et al., 1999; White et al., 1998), also called GB1 and GB2 (Fig. 1A). Heterodimerization of GB1 and GB2 is required for the formation of functional receptors, both in recombinant systems and in native tissues. Cell-surface trafficking of the $GABA_B$ receptor is controlled by an endoplasmic reticulum retention signal (RSR motif) located in the intracellular C-terminal region of GB1. This signal can be masked through a coiled-coil interaction with the intracellular tail of GB2, such that GB1 reaches the cell surface only when associated with GB2 (Calver et al., 2001; Margeta-Mitrovic et al., 2000). The $GABA_B$ receptor is an allosteric complex similar to other class C G protein-coupled receptors (GPCRs). Each subunit is composed of an extracellular domain, called Venus flytrap (VFT), which is linked to the N-terminus of a prototypical heptahelical transmembrane domain (7TM) (Pin et al., 2004). $GABA_B$ agonists and competitive antagonists (the orthosteric ligands) bind to the GB1 VFT, whereas the GB2 VFT is not bound by GABA nor, most likely, by any other ligand (Kniazeff et al., 2002). In contrast, the 7TM of GB2 is responsible for G-protein activation (Galvez et al., 2001) and contains the site of action of PAMs (Binet et al., 2004).

In order to better understand both the molecular functioning of the $GABA_B$ receptor and the mechanism of action of orthosteric and allosteric ligands, it is important to know how the $GABA_B$ VFTs dimerize and control 7TM activity. We recently demonstrated that the VFTs of the two subunits interact with each other, and also that the GB2 VFT controls agonist affinity for GB1 (Liu et al., 2004); this was recently confirmed using purified VFTs (Nomura et al., 2008). However, it is unknown whether the $GABA_B$ VFT dimer functions similarly to the VFTs of mGlu receptors (Kunishima et al., 2000; Muto et al., 2007), of ANP receptors (He et al., 2001; van den Akker, 2001) or of tyrosine kinase receptors in *Schistosoma* (Vicogne et al., 2003), to which the $GABA_B$ VFTs share equal evolutionarily distance (Fig. 1B).

Here, we have identified the VFT dimerization interface and used a glycan wedge scanning approach to analyze its functional relevance. Our data demonstrate that a direct interaction between the VFTs of the $GABA_B$ subunits is required for cell surface targeting and agonist activation of the receptor. We also provide direct evidence that a change in the dimerization interface takes place during agonist activation of the receptor.

## Results

**Bioinformatic prediction of the dimerization interface of GABA$_B$ VFTs**

Previous results have shown that GABA$_B$ VFTs form heterodimers at the surface of cells, even in the absence of the receptors' 7TM and C-terminal domains (Liu et al., 2004). To localize the dimerization interfaces of the GABA$_B$ VFTs, we first analyzed the conservation of residues at their surfaces using a set of 20 GB1 VFT sequences and 22 GB2 VFT sequences, from *Dictyostelium* to mammals (See Supplementary Fig. 1); the sequences were selected based on our previously established 3D models of the GABA$_B$ VFTs (Kniazeff et al., 2002). This analysis revealed that one face of the subunits is more conserved than the others (Face 1 in Fig. 1C and 1D), suggesting that it might correspond to the dimerization interface, as observed with mGlu receptor dimers (Rondard et al., 2006). This possibility was also supported by an analysis of the charge distribution at the surface of the VFTs. Indeed, the conserved Face 1 is composed of a large patch of hydrophobic residues in both subunits (Fig. 1C and 1D), and corresponds to the hydrophobic dimerization interface of the mGlu VFT dimer.

In view of this correspondence, the 3D structure of the mGlu VFT dimer was used to build a model of the GB1-GB2 VFT heterodimer (Fig. 2A). Interestingly, none of the putative N-glycosylation sites (consensus sequence Asn-*X*-Ser/Thr or N*X*S/T, where *X* can be any natural amino-acid except Pro) found in the GB1 and GB2 sequences from different species, from nematodes to mammals, were located within the proposed dimerization interface (Fig. 2A). In contrast, most other faces contained at least one putative glycosylation site in at least one of the species examined. This further supported our model of GB1 and GB2 VFT interaction. Taken together, these observations were consistent with the VFT dimer interface in the GABA$_B$ receptor being similar to that in mGlu receptors, involving the same two helices of lobe 1.

**Introduction of N-glycans at the VFT interface abolishes receptor activity**

In order to examine the functional importance of the interaction between the VFTs, we tried to block the interaction by introducing N-glycans at the possible dimer interface in GB1 or GB2 (Fig. 2B), creating a steric wedge. Experimentally, we introduced the consensus sequence N*X*S/T (where *X* can be any natural amino acid except Pro), which typically results in the attachment of a bulky N-glycan moiety to the side chain of the Asn residue. These N-glycosylation sites were introduced at different positions within GB1 (225, 229, 232, 251, 255, and 258) and GB2 (110, 114, 118, 137, 141 and 145) (Fig. 2B and see Supplementary Table 1). To ensure the correct trafficking of GB1 to the cell surface when expressed alone, these mutations were first introduced into a GB1 subunit that had a mutated ER retention signal (ASA instead of RSR) (Brock et al., 2005; Pagano et al., 2001).

Western blot experiments revealed that all mutated subunits were expressed at their expected molecular weights (Fig. 3A and 4A). Due to the large size of the full-length subunits (130 kDa), it was impossible to detect any shifts in their apparent molecular weights due to the presence of additional N-glycan moieties (Fig. 3A and 4A). However, such shifts were clearly seen with most mutants of GB1 and GB2 VFTs attached to the plasma membrane with a single transmembrane domain (Liu et al., 2004) (Fig. 3B and 4B, see supp Table 1). Moreover, treatment with glycosidase PNGase F restored gel mobility similar to that of wild-type VFT (Fig. 3B and 4B), demonstrating that the mutants were indeed glycosylated.

The surface expression of all the full-length glycosylation mutants was also examined using haemaglutinin (HA) or Flag epitopes inserted into the extracellular N-terminal end (after the signal peptide) of GB1$_{ASA}$ and GB2, respectively. ELISA assays performed on intact cells revealed that all mutants reached the cell surface. While a lower level was observed with some of them at the cell surface (Fig. 3C and 4C), their total expression levels as measured on western blots was similar to that of the wild-type proteins, suggesting that some mutants had difficulties passing the quality control system. This was confirmed by quantifying both the surface and total expression levels of these subunits using ELISA performed on intact and permeabilized cells, respectively (supp Fig. 3A and 3B).

The functional consequences of these additional glycosylation sites were then analyzed by co-expressing either the mutated GB1 subunits together with wild-type GB2 (Fig. 3D), or the mutated GB2 forms with $GB1_{ASA}$ (Fig. 4D). In most cases, GABA was unable to generate a response in cells co-expressing the subunits, despite a sufficient expression level of both proteins at the cell surface. Among the different mutants tested, only two GB1 mutants (HA-$GB1_{ASA}$-N225 and -N232) and one GB2 mutant (Flag-GB2-N137) were able to form a functional receptor when co-expressed with the wild-type partner. Notably, for those positions whose mutation generated non-functional receptors, it was sterically impossible to add an N-acetyl glucosamine group to the Asn residue in the 3D model of the GB1-GB2 VFT dimer, in contrast to those residues whose mutation did not interfere with the formation of a functional heterodimer (Supp. Fig. 2A and 2B).

To address whether the negative effect of the NXS/T mutations was indeed due to the introduction of an additional glycan on the VFT, we first tested the effect of the N-glycosylation inhibitor tunicamycin (Luo et al., 2003). However, this molecule was toxic to our electroporated cells, preventing us from examining its effect on receptor function. Indeed, even the response mediated by the wild-type receptor could no longer be measured. We therefore compared the properties of certain N-glycosylation site mutants (HA-$GB1_{ASA}$-N229 or -N251 and Flag-GB2-N114 or -N141) with those of analagous mutants in which the Asn residue was replaced by a Gln, which cannot be glycosylated. The Gln-containing mutants (HA-$GB1_{ASA}$-Q229 and -Q251; and Flag-GB2-Q114 and -Q141) generated similar responses upon activation with GABA to those obtained with the wild-type receptor (Fig. 5A and 5B). These data demonstrated that the lack of activity of the HA-GB1$_{ASA}$-N229 or -N251 and Flag-GB2-N114 or -N141 mutants was likely due to the presence of the N-glycan on the VFT, and not to the mutation *per se*.

**Binding and G-protein coupling properties of GB1 and GB2 mutants**

Among the mutations introduced into GB1, those that led to non-functional receptors were unable to bind the radiolabeled antagonists $^3$H-CGP54626 and $^{125}$I-CGP64132, suggesting that the mutant proteins were not folded correctly (supp Fig. 4). It has recently been reported that the correct association of soluble GB1 and GB2 VFTs is required for the GB1 VFT to be able to bind ligands (Nomura et al., 2008). Then, the absence of antagonist binding on GB1-N229 and -N251 could well be the consequence of the lack of possible association with GB2 rather than a misfolding due to the N-glycan. We therefore introduced nine additional glycosylation sites into the GB1 VFT in a region devoid of any natural sites in the various GB1 subunits identified in different species (Fig. 2). All these mutations generated GB1 VFTs with an additional N-glycan, as shown by a decrease in mobility in acrylamide gels (supp Fig. 5A). Moreover, all were expressed correctly to the cell surface, and eight led to a functional GABA$_B$ receptor upon co-expression with GB2, with GABA EC$_{50}$ values in the same range as that of the wild-type receptor (supp Fig. 5B,C). Further information on the non-functional N315 mutant is provided at the end of the results section (see Fig. 11).

Regarding the GB2 mutants, the absence of activation by GABA of the GB2-N114 and -N141 mutants was not due to the uncoupling of these subunits from G-proteins. We previously reported that the PAM of the GABA$_B$ receptor, CGP7930, has agonist activity and can activate both GB2 expressed alone and a truncated version of GB2 that is deleted for the VFT (GB2 7TM) (Binet et al., 2004). Interestingly, the GB2 subunits carrying an additional

N-glycan could still be activated by CGP7930 (Fig. 5C), demonstrating that they all retained their ability to activate G-proteins.

These data revealed that perturbation of the putative dimer interface in both the GB1 and GB2 VFTs has important consequences for the activity of the receptor, but does not prevent the 7TM of GB2 from reaching an active state.

**N-glycan in GB2 lobe 1 VFT interface prevents receptor heterodimerization**

To further analyze the molecular mechanism by which the N-glycan modification of GB2 abolishes receptor activity, we focused our study on three mutants that were well-expressed (N114, N137 and N141), but of which only one, GB2-N137, forms a functional GABA$_B$ receptor when co-expressed with GB1$_{ASA}$. Using FRET measurements, we found that GB2-N114 and -N141 do not interact with GB1 at the cell surface, in contrast to GB2-N137. These experiments were conducted on intact cells, using anti-HA antibodies conjugated with the energy donor fluorophore (europium cryptate PBP) to label the HA-tagged GB1, and anti-Flag antibodies linked to the fluorophore acceptor (d2) to label the Flag-tagged GB2 proteins (Fig. 6A). Such an approach enables the detection of receptor dimers at the cell surface only, as previously shown with GB1 and GB2 subunits co-expressed in the same cells (Maurel et al., 2004). As shown in Fig. 6B, a very low FRET signal was measured between Flag-GB2-N114 or -N141 and HA-GB1$_{ASA}$ that was not significantly different from a negative control (FRET between Flag-GB2 and HA-CD4 for example, data not shown). In contrast, large FRET signals were obtained between HA-GB1$_{ASA}$ and either Flag-GB2-N137, -Q114, -Q141 or wild-type Flag-GB2 (Fig. 6B). These experiments were conducted with similar amounts of GB1$_{ASA}$ and GB2 subunits at the cell surface for all the constructs tested, to ensure that differences in FRET signals did not reflect differences in the level of expression of one of the subunits (Fig. 6C and 6D). These data strongly suggest that the glycosylation site prevents the direct interaction of GB1 with GB2. Co-immunoprecipitation experiments confirmed that, among the GB2 mutants, only GB2-N137 interacts with GB1 (supp Fig. 6).

One well-established consequence of the GB1-GB2 interaction is an increase in agonist affinity for GB1. Indeed, GABA affinity, as measured by the displacement of $^{125}$I-CGP64213 on GB1$_{ASA}$ expressed alone, was 5-10 fold lower than the affinity measured in the presence of GB2 (Liu et al., 2004) (Fig. 7). Co-expression of GB1$_{ASA}$ with GB2-N114 or -N141 did not increase agonist affinity for GB1$_{ASA}$ (Fig. 7), whereas GB2-N137, -Q114 and -Q141 had the same effect as wild-type GB2. These data indicate that the presence of a glycosylation site at the GB2 VFT dimer interface suppresses the allosteric control of agonist affinity in GB1, consistent with a lack of interaction between GB2-N114 and -N141 and GB1 (Fig. 6B).

**The presence of N-glycan at the GB2 lobe 1 VFT interface abolishes cell surface targeting of the heterodimer**

Wild-type GB1 is retained in the endoplasmic reticulum in the absence of GB2 (Calver et al., 2001; Margeta-Mitrovic et al., 2000; Pagano et al., 2001). Up to this point, all of the experiments had been conducted with a mutated version of GB1 in which the ER retention signal was changed to ASA. Therefore, we next examined if the glycosylated GB2 mutants could still target wild-type GB1 to the cell surface. Indeed, the inactive N-glycan-modified GB2-N114, -N118 and -N141 receptors were unable to target wild-type GB1 to the cell surface, as indicated by both ELISA and $^{125}$I-CGP64213 binding experiments on intact cells (Fig. 8). In contrast, GB2-Q141 was able to target GB1 to the cell surface. These results indicate that the masking of the ER retention signal cannot occur if the VFT interaction is prevented by the presence of an N-glycan, even at an early stage of protein processing. However, a GB2 deletion mutant lacking the VFT was still able to target GB1 to the cell surface (Fig. 7), as can a GB2 coiled-coil domain expressed alone (Brock et al., 2005). Taken together, these data demonstrate that: i) the coiled-coil interaction between the $GABA_B$ subunits plays a major role in controlling the trafficking of GB1 to the cell surface; and ii) this interaction can only occur if subunits are able to interact normally outside of the coiled-coil domain. In other words, the interaction of the coiled-coil domains within the full-length subunits can only take place if the VFTs are able to interact.

**N-glycan at the GB1 lobe 1 interface abolishes the interaction with GB2**

One possible explanation for the observed absence of antagonist binding in the GB1 mutants carrying an N-glycan at the dimerization interface was that they were unable to associate with partner VFTs, as reported with the purified soluble VFTs (Nomura et al., 2008). As observed with the GB2 mutants, low FRET signals were obtained at the cell surface between Flag-GB2 and HA-GB1$_{ASA}$-N229 or –N251, whereas significant FRET signals were measured with the functional GB1$_{ASA}$ mutants HA-GB1-N225, -Q229 and -Q251 (supp Fig. 7). Moreover, among these GB1 mutants carrying the intracellular retention signal, only the –N225 and –Q251 could reach the cell surface when co-expressed with GB2 (Supp Fig. 8). In contrast, all GB1 mutants reached the cell surface when co-expressed with a GB2 deleted for its VFT (GB2-7TM, data not shown).

**Introduction of an N-glycan into the VFT lobe 2 interface locks the receptor in an inactive state**

According to the structure of the isolated dimeric mGlu VFTs determined in the presence and absence of agonist, the lobe 2 domains within the VFT dimer are far apart in the inactive state (Kunishima et al., 2000) but are in contact in the active receptor, constituting a $Gd^{3+}$ binding site (Tsuchiya et al., 2002). Such a major movement of the mGlu VFTs relative to one another is assumed to play a critical role in the activation of these dimeric receptors, although this has never been firmly demonstrated with the native receptor dimer. We investigated whether the GABA$_B$ VFTs show a similar movement during receptor activation. To do this, we introduced an N-glycan wedge into the lobe 2 of GB2, at position 209 (Flag-GB2-N209), in the region corresponding to the Gd$^{3+}$ binding site in mGluR1 (Fig. 9A). Interestingly, this GB2 mutant produced an inactive GABA$_B$ receptor when co-expressed with GB1, while the non-glycosylated version of this mutant (Flag-GB2-Q209) gave rise to a perfectly functional receptor (Fig. 9B and C). This loss of activity was due to neither impaired expression of Flag-GB2-N209 at the cell surface (Fig 9D), nor to a defect in assembly with GB1, as shown by: *i*) the correct cell-surface targeting of wild-type GB1 (Fig. 10A); *ii*) a positive allosteric effect of GB2-N209 on the affinity of GABA for GB1 (Fig. 10B); and *iii*) a FRET signal similar to that measured with the wild-type receptor (Fig. 10C). The lack of activity of the GB1:Flag-GB2-N209 dimer was also not due to misfolding of the GB2 subunit, since this receptor combination could still be activated by the PAM CGP7930 (Fig. 10D).

Similar data were obtained with an equivalent mutant at the GB1 lobe 2 interface, GB1-N315 (Fig.11). Indeed, this GB1 mutant co-expressed with GB2 is not functional (Fig. 11B), although it is well expressed and targeted to the surface in the presence of GB2 (Fig. 11C, E), glycosylated (Fig. 11D) and still able to bind CGP64213 with wild-type affinity (Fig. 11F).

These data revealed that the presence of an N-glycan wedge in the lobe 2 interface prevents the heterodimer from reaching the active state upon agonist binding in GB1, but does not prevent the correct assembly of the receptor heterodimer.

## Discussion

How the signal is transduced in the $GABA_B$ heterodimer, from the agonist-occupied GB1 VFT to the GB2 7TM responsible for G-protein activation, remains unknown. In the present study, we have used a glycan wedge scanning approach to identify the dimerization interface within the VFT dimer. We also show that the physical interaction between the subunits at the interface is necessary for receptor function, for allosteric interaction between the subunits, and for the correct assembly and trafficking of the $GABA_B$ heterodimer to the cell surface. We also provide evidence that a change in relative position between the VFTs is required for signal transduction to occur.

**The glycan wedge approach**

Random insertion of glycosylation sites has often been used to study the topology of transmembrane proteins, for instance to identify their extracellular portions. However, many mutants are not functional and are likely misfolded due to the introduction of N-glycans in buried positions (see for example (Hollmann et al., 1994)). However, if the 3D protein structure is taken into account, and the Asn side chain is well exposed at the surface of the protein, then the N-glycan is not expected to dramatically affect the folding the protein. Indeed, natural glycosylation sites are typically found in loops, helices, and beta strands. Significantly, among the 13 glycosylation sites introduced into the GB1 VFT, 11 resulted in the correct folding of the domain, as illustrated by their exhibiting normal activity or binding.

The insertion of N-glycans at critical places can be used to prevent protein-protein interactions involving extracellular domains, or to block specific conformational changes. Although this has never been used systematically, the insertion of an N-glycan was used to block the activation of the β-integrin receptor (Luo et al., 2003). In addition, a natural mutation creating a glycosylation site has been identified at the level of the dimerization interface of the sweet and umami taste receptor subunit T1R3, leading to non-tasting mice (Max et al., 2001). It was therefore speculated that the additional N-glycan prevented the association of the taste receptor subunits, thus preventing their normal functioning. This is consistent with our finding that glycosylation at the same place in either GB1 or GB2 prevents the association of these subunits, and consequently the functioning of the $GABA_B$ receptor.

**Conservation of the VFT dimerization interface**

Despite the evolutionary distance of the $GABA_B$ receptor VFTs from those of other class C GPCRs, we provide evidence that their dimerization interface is conserved. This is nicely illustrated by the high degree of conservation of this interface between the mGlu and $GABA_B$ VFTs, by the hydrophobicity of the $GABA_B$ interface, and, most importantly, by the demonstration that the presence of an N-glycan at this interface in either GB1 or GB2 prevents the interaction between the subunits. It may appear surprising that the insertion of N-glycan at this putative dimer interface in GB1 results in incorrectly folded mutant subunits, as shown by their inability to bind $GABA_B$ antagonists. Although one cannot exclude the possibility that the misfolding results from the N-glycan insertion *per se*, thus preventing dimer formation, the misfolding may simply result from a lack of interaction with the partner VFT at this level. Indeed, such a hydrophobic dimerization area may not be stable when exposed to aqueous solvent. In agreement with this hypothesis, it has recently been shown that the soluble GB1 VFT does not fold correctly in the absence of its partner GB2 VFT (Nomura et al., 2008); this provides further evidence that the region identified corresponds to the dimerization interface of the $GABA_B$ VFTs.

Interestingly, the same area also serves as a dimerization interface in other VFT-containing proteins, such as the atrial natriuretic peptide receptors (He et al., 2001; van den Akker, 2001), as well as in prokaryotic VFTs (Schumacher et al., 1994). This suggests that this mode of association between the VFTs arose early in evolution, likely before the appearance of class C GPCRs.

**Conformational change of the GB1-GB2 VFTs interface during activation**

The current hypothesis for the ligand-induced activation of class C GPCRs is that a change in the relative position of the VFTs leads to a relative movement of the 7TMs and, as a result, the activation of one of them (Pin et al., 2005). The most compelling evidence supporting this idea comes from the crystal structure of the soluble dimeric mGlu1 VFTs (Kunishima et al., 2000; Tsuchiya et al., 2002). However, it remains to be proven that what was observed with the soluble mGlu1 VFTs is related to the activation process of the full-length receptor, and also that the findings would apply to the structurally distant $GABA_B$ receptor. Indeed, the specific lobe 2 interaction observed in the glutamate bound form of the mGlu1 VFTs (Tsuchiya et al., 2002) is not observed in the soluble dimeric mGlu3 VFTs occupied with five different agonists (Muto et al., 2007) in which the lobes 2 are still apart.

Our data show that the insertion of an N-glycan at the lobe 2 interface does not prevent the two $GABA_B$ VFTs from interacting, nor does it affect the cell surface targeting of the heterodimer. In addition, and most importantly, the N-glycan does not impair the ability of GB2 to increase agonist affinity on GB1, indicating that the two subunits are correctly assembled. This finding indicates that an N-glycan at that position can well be accommodated in the inactive VFT dimer structure, in agreement with the idea that the lobes 2 do not interact with each other in the inactive receptor. However, the presence of this N-glycan does prevent agonist activation of the receptor, suggesting that N-glycan at this site interferes with the adoption of the active form of the VFT dimer. This is entirely consistent with the notion that there is a relative movement of the VFTs upon agonist binding, similar to what is observed in the crystal structure of the dimeric mGlu1 VFT. These data not only support the view that a relative movement of the GABA$_B$ VFTs is required for agonist activation, but also that this may be a general mechanism among class C GPCRs in spite of their structural differences.

Surprisingly, even though the GB2-N209 mutant does not allow agonist activation, it still enhances agonist affinity on GB1, indicating that this increased affinity is unrelated to the active state conformation. This is consistent with our previous results showing that this effect of GB2 mostly occurs indirectly, alleviating an inhibition of the GB1 VFT by the GB1 7TM as a result of the interaction between the VFTs, rather than from a direct allosteric action between the VFTs (Liu et al., 2004).

We also found that although agonists were no longer able to activate the GB2-N209-containing receptor, the positive allosteric modulator CGP7930 could. This shows that the effect of this molecule, which acts upon the GB2 7TM, does not require the relative movement of the VFTs.

**Importance of VFTs for GABA$_B$ receptor assembly**

Our results provide new insight into how GABA$_B$ subunits assemble together to form a functional heterodimer at the cell surface. One limiting step requires the masking of the ER retention signal in GB1 through a coil-coiled interaction involving the intracellular C-terminal portions of the two subunits (Brock et al., 2005; Calver et al., 2001; Margeta-Mitrovic et al., 2000; Pagano et al., 2001). Indeed, we previously reported that the GB2 CC domain is sufficient for masking the GB1 ER retention signal (Brock et al., 2005), as is a GB2 subunit deleted for its VFT (this study), indicating that the interaction between the VFTs is not required *per se* for this process. However, preventing the interaction between the VFTs is sufficient to prevent the masking of the GB1 ER retention signal, indicating that the CC interaction cannot occur if the two VFTs of the full-length subunits cannot assemble correctly. This indicates that the VFT interaction plays a crucial role in the correct assembly of the functional GABA$_B$ receptor. Such a role for VFT dimerization in the assembly of oligomeric proteins has also been reported with ionotropic glutamate receptors (Ayalon et al., 2005; Leuschner and Hoch, 1999; Meddows et al., 2001).

Taken together, our data, obtained using an N-glycan wedge scanning approach, has revealed the critical importance of the correct association of the $GABA_B$ VFTs for receptor function and for passing the quality-control checkpoints in the biosynthetic pathways. They have also shed new light on the activation process of this prototypical GPCR dimer, revealing the requirement for a relative movement between the VFTs for the action of agonists, but not for that of positive allosteric modulators.

## Materials and methods

**Materials**

GABA was obtained from Sigma (Missouri, USA). CGP7930 and [$^{125}$I]-CGP64213 were purchased from Tocris (Fisher Scientific BioBlock, Illkrich, France) and Anawa (Zurich, Switzerland), respectively.

**Plasmids and transfection**

pRK5 plasmids encoding the wild-type GB1a, GB1$_{ASA}$ and GB2 subunits tagged with a hemagglutinin (HA) or Flag epitope at their N-terminal end under the control of a cytomegalovirus promoter, were described previously (Galvez et al., 2001; Pagano et al., 2001). GB1 and GB2 mutants were generated using the QuikChange mutagenesis protocol (Stratagene, La Jolla, CA). For western blot analysis of N-glycan introduced by the mutations, EcoRI-BamHI fragments of pRK-HA-GB1 and pRK-Flag-GB2 containing the mutations were subcloned into pRK vectors encoding the GB1 and GB2 subunits deleted for the transmembrane and C-terminal tail, respectively, called ΔGB1 and ΔGB2 (Liu et al., 2004).

HEK-293 and COS-7 cells were cultured in Dulbecco's modified Eagle's medium supplemented with 10% FBS and transfected by electroporation as described elsewhere (Liu et al., 2004). Ten million cells were transfected with plasmid DNA containing GB1$_{WT}$ (2 µg), GB1$_{ASA}$ (2 µg), GB2$_{WT}$ (2 µg), or one of the mutants (6 µg), and completed to a total amount of 10 µg plasmid DNA with empty pRK5 vector. To allow efficient coupling of the receptor to the phospholipase C pathway, the cells were also transfected with the chimeric Gαqi9 (2 µg) (Galvez et al., 2001).

**Deglycosylation assays**

Twenty hours after transfection, COS-7 cells were washed with PBS (Ca$^{2+}$- and Mg$^{2+}$-free) and harvested. The membranes were prepared as previously described (Liu et al., 2004).

For each sample, 50 µg of total protein was denatured using 0.5% (w/v) SDS, 1% Nonidet P-40, 1% (v/v) β-mercaptoethanol, Tris HCl pH 7.4, 50 mM NaCl, protease inhibitor cocktail, for 5 min at 50°C, and incubated with 1 U of N-glycosidase F (Roche, Penzberg, Germany) for 2 h at 37°C. Reactions were stopped by adding 6 x sample buffer and heating the samples at 95°C, before western blotting.

**Co-immunoprecipitation assays**

As previously described (Michineau et al., 2006; Terrillon et al., 2003), twenty hours after transfection, COS-7 cells were washed with PBS ($Ca^{2+}$- and $Mg^{2+}$- free), and incubated for 1 h in blocking buffer (PBS containing 0.2% BSA) and for 2 h with a monoclonal rat anti-HA antibody 3F10 (Roche) at 4°C. After two washes in blocking buffer, cells were lysed 45 min at 4°C in RIPA buffer (50 mM Tris HCl pH 7.4, 150 mM NaCl, 1% Nonidet P-40, 0.5% sodium deoxycholate, 0.1% SDS, protease inhibitor cocktail), and centrifuged at 12,000 x g for 30 min at 4°C. Lysates were incubated with immobilized Protein A/G (Pierce, Rockford, IL) beads for 3 h at 4°C. After three washes with PBS, the beads was resuspended in 2 x sample buffer, and heated at 95°C, before western blotting.

**Western blotting**

For each sample, 50 µg of total protein was subjected to SDS-PAGE using 10% polyacrylamide gels. Proteins were transferred to nitrocellulose membrane (Hybond-C; Amersham Biosciences). HA-tagged proteins were probed with a polyclonal anti-HA rabbit antibody (dilution 1/400; Zymed, San Francisco, CA) and then with an Alexa-Fluor® 700 goat anti-rabbit antibody (dilution 1/3000; Molecular Probes Invitrogen Corporation, Carlsbad, CA). Flag-tagged constructs were probed with the mouse monoclonal anti-Flag antibody M2 (Sigma, St. Louis, MO) at 0.6 µg/mL, and then with a DyLight™ 800 anti-mouse antibody (Rockman Immunochemicals, Inc., Gilbertsville, PA) at 0.1 µg/mL. Proteins were visualized by the Odyssey® Infrared Imaging System (Li-Cor Biosciences, Lincoln, NE).

**Inositol phosphate and intracellular calcium measurements**

Measurements of inositol phosphate (IP) accumulation and the calcium signal in transfected HEK-293 cells were performed in 96-well microplates as previously described (Goudet et al., 2004).

**Cell surface quantification by ELISA and ligand binding assays**

Experiments were conducted as described (Liu et al., 2004). HA-tagged constructs were detected with a monoclonal rat anti-HA antibody 3F10 (Roche) at 0.5 µg/mL and goat anti-rat antibodies coupled to horseradish peroxidase (Jackson Immunoresearch, West Grove, PA) at 1.0 µg/mL. Flag-tagged constructs were detected with the mouse monoclonal anti-Flag antibody M2 (Sigma, St. Louis, MO) at 0.8 µg/mL and goat anti-mouse antibodies coupled to horseradish peroxidase (Amersham Biosciences, Uppsala, Sweden) at 0.25 µg/mL.

The ligand binding assay on intact HEK-293 cells was performed as previously described using 0.1 nM [$^{125}$I]-CGP64213 (Anawa, Zurich, Switzerland) (Liu et al., 2004). The radioligand was displaced by increasing concentrations of GABA (Sigma, Saint Louis, MO). The curves were fitted according to the equation: "$y=[(y_{max}-y_{min})/(1+(x/IC_{50})^{nH})] +y_{min}$" using GraphPad Prism software (San Diego), where the $IC_{50}$ is the concentration of the compound that inhibits 50% of bound radioligand and nH is the Hill coefficient.

**Time-resolved FRET measurements**

Time-resolved FRET experiments were conducted as described (Maurel et al., 2004). This methodology is based on the non-radiative energy transfer between rare earth cryptates such as europium ($Eu^{3+}$) cryptates and the acceptor fluorophore d2 (CisBio International). Briefly, COS-7 cells co-expressing the HA-tagged GB1 and Flag-tagged GB2 were incubated with 1 nM of monoclonal anti-HA (12CA5) antibodies carrying $Eu^{3+}$-Cryptate PBP and 3 nM of monoclonal anti-Flag (M2) d2 (provided by Cis Bio International, Bagnols-sur-Cèze, France). As a negative control, cells were incubated with the fluorescence donor antibodies only.

**Molecular modelling**

A homology model of the $GABA_B$ VFTs was generated using the crystal structure of mGlu1 VFT (Protein Data Bank accession number 1EWK) as a template. Models were manually refined with ViTO (Catherinot and Labesse, 2004) using the sequence alignment of the $GABA_B$ VFTs. Final models were built using Modeller 7.0 (Sali and Blundell, 1993) and evaluated using the dynamic evolutionary trace as implemented in ViTO.

## Acknowledgments

We would like to thank Dr. P. Paoletti for constructive discussions and N. Liu, J. He and X. Li for initial work. The authors wish to thank F. Maurin (CisBio International) for the preparation of the labeled antibodies, and C. Vol for the support with the $Ca^{2+}$ assay. This work was made possible thanks to the screening facilities of the Institut Fédératif de Recherche (IFR) 3. This work was supported by a CNRS/NSFC collaboration project (PICS3604) (to P.R. and J.L.), the National Basic Research Program of China (Grant 2007CB914202 to J. L.), the National Natural Science Foundation of China (Grants 30530820 and 30470368 to J. L.), and the Hi-Tech Research and Development Program of China (863 project) (Grant 2006AA02Z326 and 2007AA02Z322 to J. L.). J.-P. P. was supported by the Centre National de la Recherche Scientifique (CNRS), the Institut National de la Santé et de la Recherche Médicale (INSERM), and by grants from the French Ministry of Research, Action Concertée Incitative "Biologie Cellulaire Moléculaire et Structurale" (ACI-BCMS 328), the Agence Nationale de la Recherche (ANR-BLAN06-3_135092), and by an unrestricted grant from Senomyx (La Jolla, CA, USA).

**Figure legends**

**Figure 1**

**Models and bioinformatic analysis of the GABA$_B$ VFTs. A.** Structural model of the dimeric GABA$_B$ receptor in the resting state. Corey-Pauling-Koltun representation of the GABA$_B$ VFT dimer model generated according to the resting state of the dimeric mGlu receptors (PDB accession number 1EWT), and apposition of two heptahelical domains (7TM) according to the rhodopsin dimer structure (PDB accession number 1n3m). GB1 (*yellow*) and GB2 (*blue*) are in the front and the back, respectively. The C-terminal regions of the two subunits are associated through a coil-coiled (CC) interaction that masks the RSR intracellular retention signal of GB1. **B.** The phylogenetic tree was constructed using the sequences of the VFTs of the mGlu1 receptor, the amide-binding protein (AmiC) from the amidase operon, the NR2A subunit of the rat N-methyl-D-aspartate (NMDA) receptor, the leucine-isoleucine-valine-binding protein (LIVBP), the natriuretic peptide receptors types A and C (NPRA and NPRC, respectively), RTK1 from Schistosoma mansoni, and the rat GB1 and GB2 subunits. Only branches with bootstrap values >600 are shown. **C and D**, Evolutionary conservation of residues (*upper panels*) and electrostatic surfaces (*lower panels*) of the GB1 and GB2 VFTs visualized on both faces of the VFTs (Face 1 and Face 2). Conservation scores are indicated according to a color scale, from variable (*blue*) to conserved (*purple*) residues. No conservation scores were calculated for the residues *in grey*. Electrostatic surface representations are provided (negative, *red*; neutral, *white*; positive, *blue*) of the VFT faces in which the *green ribbons* correspond to the helices of the associated subunit in the inactive state, illustrating the possible dimerization interface.

**Figure 2**

**Native and engineered N-glycan sites in the heterodimeric GABA$_B$ VFTs. A**. Ribbon views of the heterodimeric VFTs are shown, with the putative N-glycosylation sites (Cα of Asn residue) in mammalian VFTs are *in cyan* and *orange* for GB1 and GB2, respectively. Additional putative N-glycosylation sites in other species are *in dark blue* and *magenta* for GB1 and GB2, respectively. **B.** The GABA$_B$ VFT interface is mainly composed of two helices (*green*) in lobe 1 of GB1 and GB2 that interact together. The positions of Cα of Asn residues modified by a, N-glycan, and resulting in a non-functional or functional receptor, are depicted *in red* and *blue*, respectively.

**Figure 3**

**Analysis of the N-glycosylated GB1 mutants. A**. Western blot analysis of the full-length HA-GB1 mutants from membrane fractions of cells co-expressing HA-GB1 mutants and Flag-GB2-WT. Cartoons depicted the GABA$_B$ subunits wild-type (*white*) and mutant (*grey*), and engineered N-glycan is indicated by a white star. **B**. Western blot analysis of the truncated GB1 subunits deleted of both HD and C-terminal regions, with or without treatment with PNGase F, and comparison to the wild-type construct. **C.** Amount of HA-tagged GB1 mutants co-expressed with Flag-tagged GB2-WT at the cell surface as measured by ELISA. **D.** IP production for the HA-tagged GB1$_{ASA}$ mutants co-expressed with GB2-WT. Data are means ± S.E. of at least three independent determinations.

**Figure 4**

**Analysis of the N-glycosylated GB2 mutants. A**. Western blot analysis of the full-length Flag-GB2 mutants from membrane fractions of cells co-expressing Flag-GB2 mutants and HA-GB1$_{ASA}$. As indicated in Figure 3, cartoons depicted the GABA$_B$ subunits wild-type (*white*) and mutant (*grey*), and engineered N-glycan is indicated by a white star. **B**. Western blot analysis of the truncated GB2 subunits deleted of both HD and C-terminal regions, with or without treatment with PNGase F, and comparison to the wild-type construct. **C.** Amount of Flag- GB2-WT mutants co-expressed with HA-GB1$_{ASA}$ at the cell surface as measured by ELISA. **D.** IP production for the Flag-tagged GB2 mutants co-expressed with $GB1_{ASA}$. Data are means ± S.E. of at least three independent determinations.

**Figure 5**

**Loss of function is due to the N-glycan at the interface between the VFTs. A.** Comparison of the IP production stimulated by the $GB1_{ASA}$ NXS/T and QXS/T mutants. **B**. Similar comparison of IP stimulation for the GB2 mutants. Data are means ± S.E. of at least three independent measurements. **C**. Effect of CGP7930 on $Ca^{2+}$ signals in cells expressing the indicated GB2 mutants. Data are means ± S.E. of triplicates from a typical experiment.

**Figure 6**

**N-glycan at the GB2 VFT interface prevents dimerization with GB1. A.** The schemes depict the experimental approach used to monitor receptor dimers at the cell surface using time-resolved FRET. The FRET signal is measured between an anti-HA antibody linked to a donor molecule (*D*) and an anti-Flag antibody linked to an acceptor molecule (*A*). **B**. FRET signal between HA-tagged $GB1_{ASA}$ and Flag-tagged GB2 subunits in cells co-expressing the indicated constructs. **C and D.** Amount of HA- and Flag-tagged subunits expressed at the cell surface, respectively, as measured by ELISA. Data are means ± S.E. of triplicates from a typical experiment.

**Figure 7**

**N-glycan at the GB2 VFT interface prevents allosteric interaction with GB1. A.** Displacement of non-permeant antagonist [$^{125}$I]-CGP64213 by GABA on HA-tagged $GB1_{ASA}$, expressed alone or in combination with the indicated Flag-tagged GB2 subunits. **B.** Amount of Flag-tagged GB2 subunits expressed at the cell surface as measured by ELISA in the same experiment. Data are means ± S.E. of triplicates from a typical experiment.

**Figure 8**

**N-glycan at the GB2 VFT interface abolishes cell surface targeting of the heterodimer.** Cell surface targeting of HA-tagged GB1-WT, expressed alone or in combination with the indicated Flag-tagged GB2 mutants and measured either by antagonist binding to GB1 (*left panel*) or anti-HA ELISA (*right panel*). Data are means ± S.E. of triplicates from a typical experiment.

**Figure 9**

**N-glycan at the GB2 VFT lobe 2 interface locks the receptor in the inactive state. A.** Structural model of the dimeric $GABA_B$ receptor based on the resting state of the dimeric mGlu1 receptors (Roo, PDB accession number 1EWT), with both VFTs open, and on the active state of mGlu1 (Aco, PDB accession number 1EWK), with the VFT of GB1 and GB2 closed and open, respectively. Note the position of residue 209 (*red*) in lobe 2 of GB2, where the engineered N-glycan prevents the receptor from reaching the active state, and the C-terminal ends of the VFTs (*green*). **B.** Intracellular $Ca^{2+}$ response mediated by the indicated wild-type and mutant Flag-tagged GB2, co-expressed with HA-tagged GB1-WT. **C.** Western blot analysis of the Flag-GB2 mutants in truncated subunits, as in Fig. 3A. **D.** Amount of Flag-tagged GB2 expressed at the cell surface as measured by ELISA.

**Figure 10**

**Loss of activity of the GB2-N209 subunit is not due to a defect in assembly with GB1. A.** Cell surface targeting of HA-tagged GB1-WT by Flag-tagged GB2 mutants as indicated, and measured by anti-HA ELISA. **B.** Displacement of non-permeant antagonist [$^{125}$I]-CGP64213 by GABA on HA-tagged GB1$_{ASA}$, expressed alone or in combination with the indicated GB2 subunits. **C.** FRET signal between HA-tagged GB1$_{ASA}$ and Flag-tagged GB2 subunits in cells co-expressing the indicated constructs. **D.** Effect of CGP7930 on $Ca^{2+}$ signals in cells expressing the indicated GB2 subunit mutants. Data are means ± S.E. of triplicates from a typical experiment.

**Figure 11**

**N-glycan in GB1 VFT lobe 2 interface locks the receptor in the inactive state. A.** Structural model of the dimeric GABA$_B$ receptor as shown in Fig. 9. Note the position of residue 315 (*red*) in the lobe 2 of GB1 where engineered N-glycan prevents the receptor activation to reach the active state, and the C-terminal ends of the VFTs (*green*). **B.** Intracellular Ca$^{2+}$ response mediated by the indicated wild-type and mutant HA-tagged GB1 co-expressed with Flag-tagged GB2-WT. **C.** Amount of HA-tagged GB1 detected at the cell surface (non-perm) or when the cells are permeabilized (perm), as measured by ELISA. **D.** Western blot analysis of the HA-GB1 mutants in truncated subunit as in Fig. 3B. **E.** Effect of CGP7930 on Ca$^{2+}$ signals in cells expressing the indicated GB1 subunits mutants. **F.** Displacement of non-permeant antagonist [$^{125}$I]-CGP64213 binding by GABA on HA-tagged GB1 mutants co-expressed with GB2-WT.

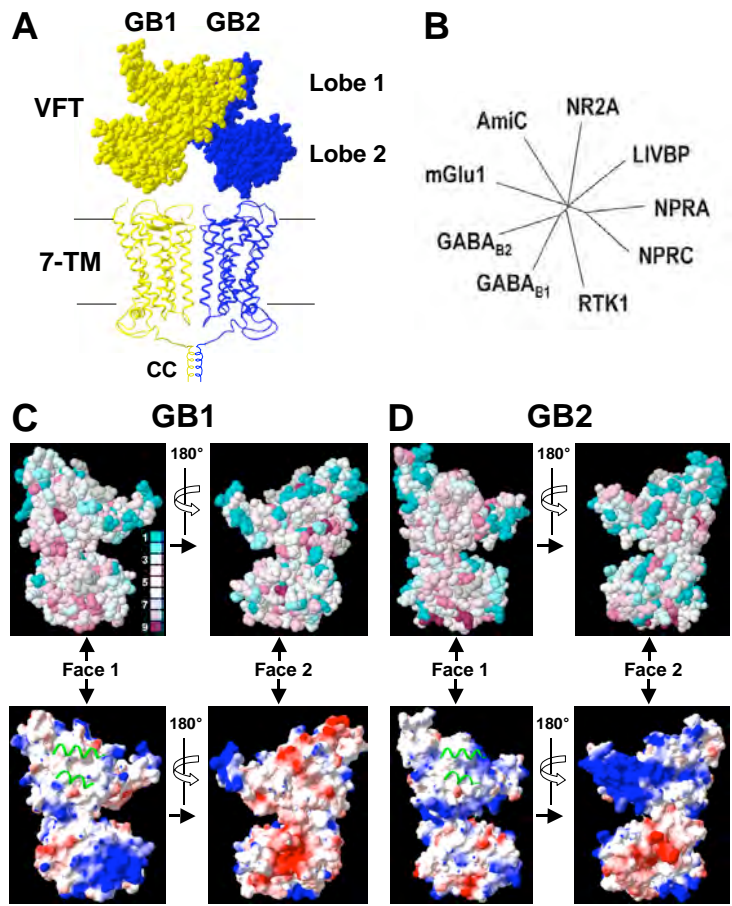

Figure 1

Figure 2

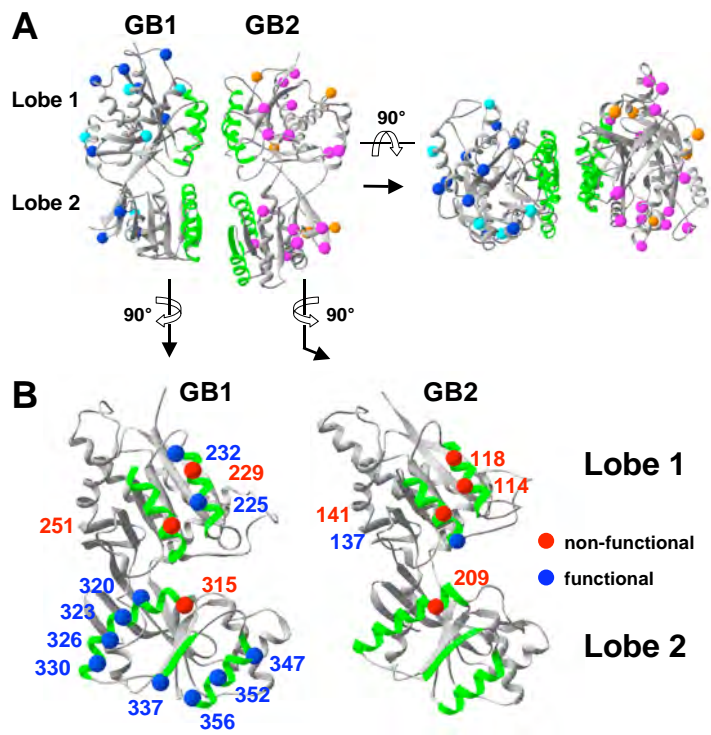

Figure 3

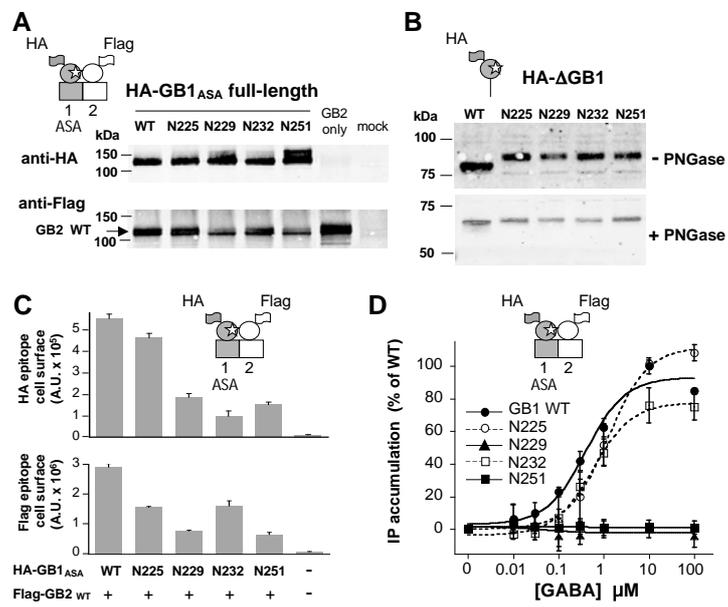

Figure 4

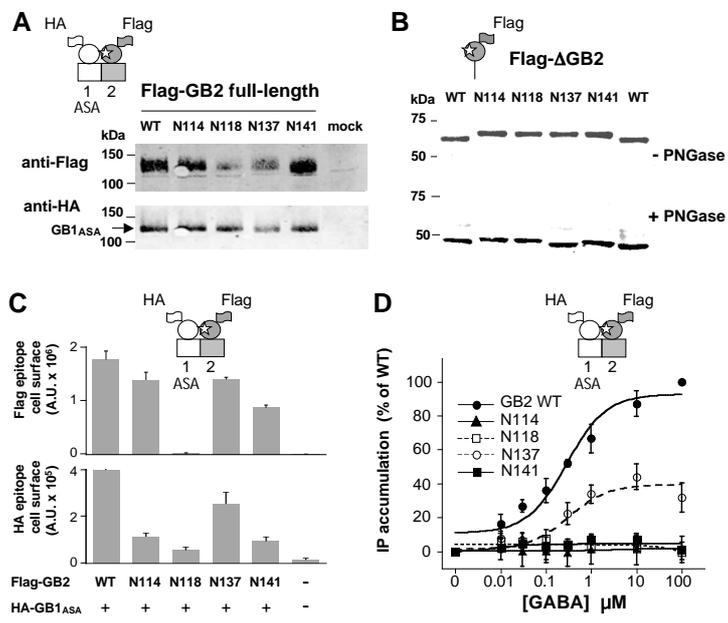

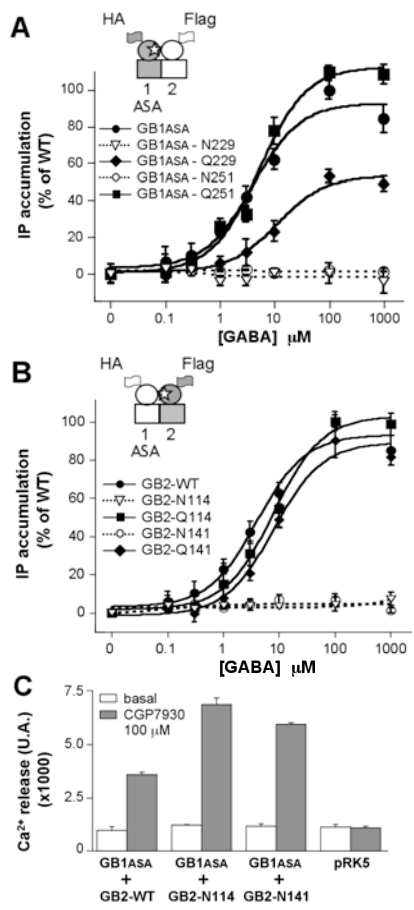

Figure 5

Figure 6

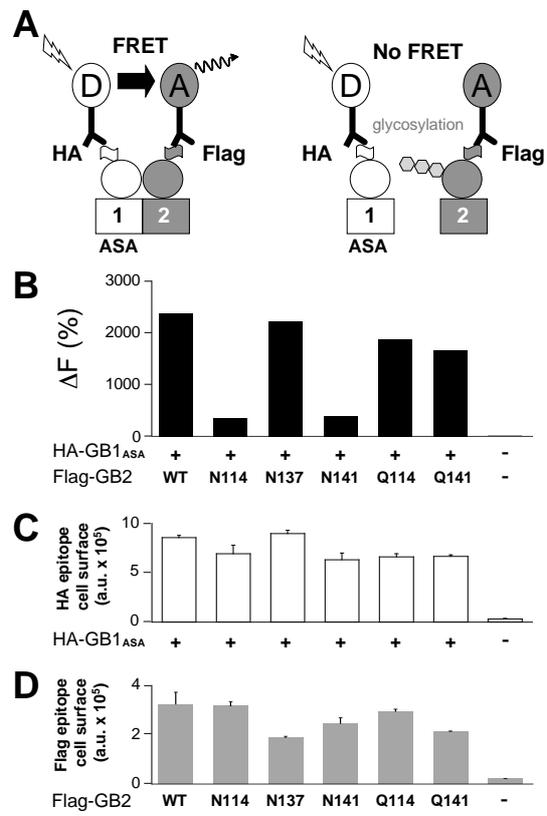

Figure 7

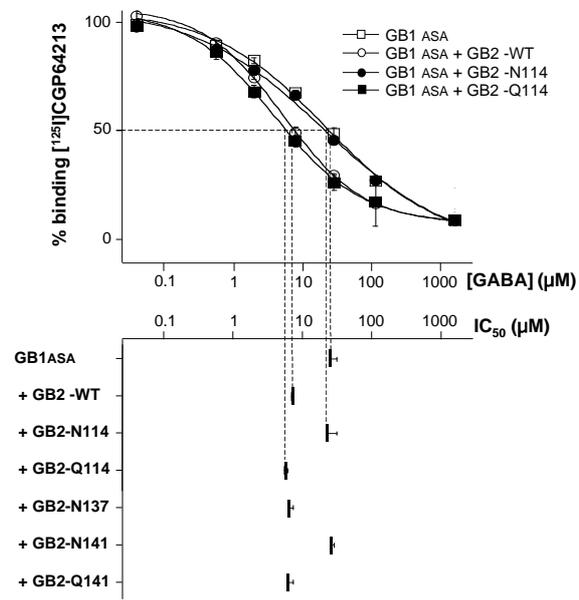

Figure 8

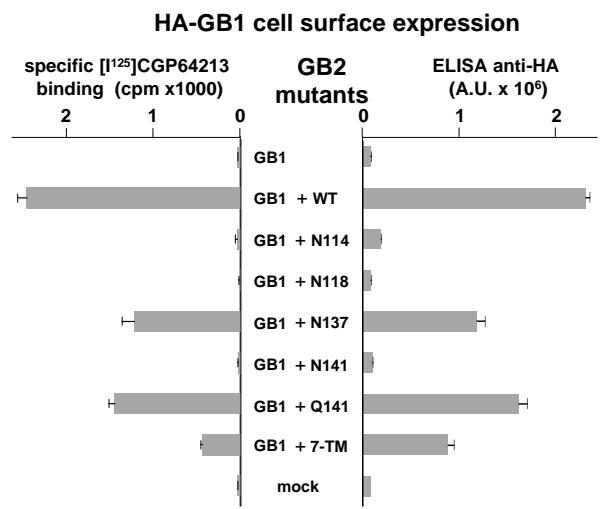

Figure 9

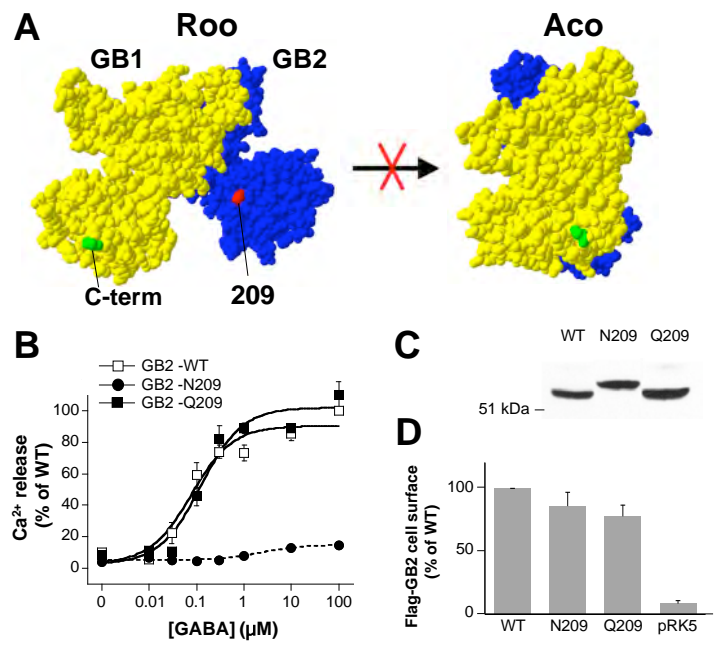

Figure 10

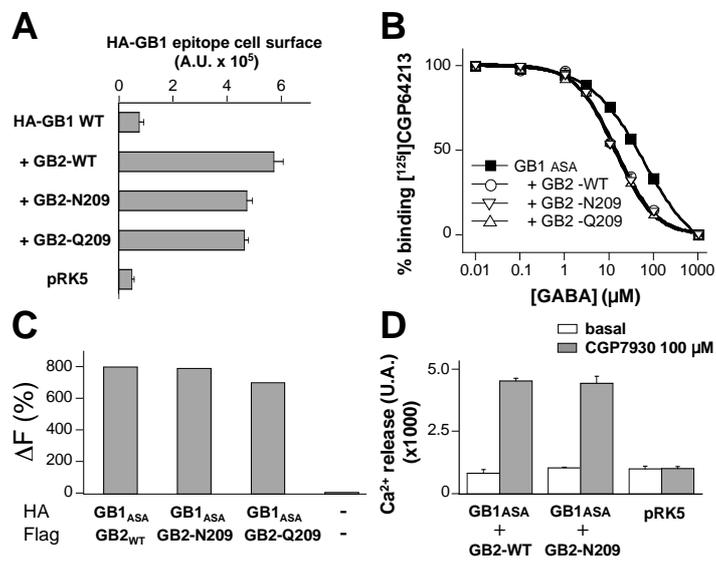

Figure 11

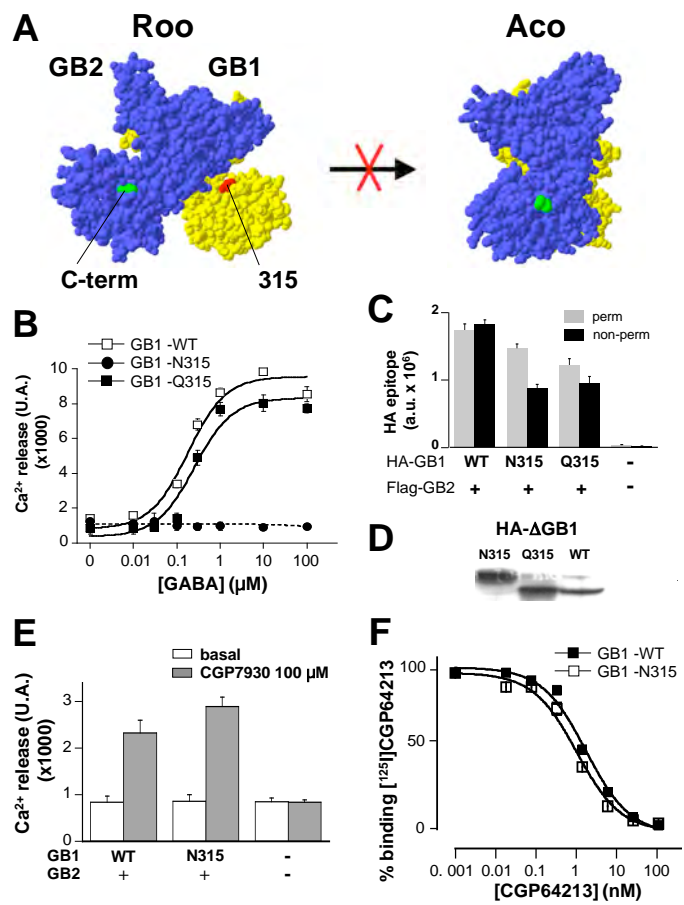